\documentclass{iaus}
\usepackage{graphicx}

\title[Dynamical modeling and the interactions with the ISM] 
{Dynamical modeling and the interactions with the ISM}

\author[Wolfgang Steffen]   
{Wolfgang Steffen$^1$
  \thanks{Present address: Institut f\"ur Computergraphik, Technische Universit\"at Braunschweig, M\"uhlenpfordstr. 23,  38118 Braunschweig, Germany}}

\affiliation{$^1$Instituto de Astronom\'{\i}a, Universidad Nacional Aut\'onoma de M\'exico, 
 Ensenada, Mexico
 \\ email: {\tt wsteffen@astrosen.unam.mx} }

\pubyear{2012}
\volume{xxx}  
\pagerange{1--8}
\jname{Title of your IAU Symposium}
\editors{A.C. Editor, B.D. Editor \& C.E. Editor, eds.}
\begin{document}

\maketitle

\begin{abstract}
This paper is a review of some of the recent modeling efforts to improve our understanding of 
structure formation and evolution of planetary nebulae including their interaction
with the interstellar medium. New propositions have been made for the formation mechanism
of multi-polar PNe and PPNe. These mechanisms are based on the central engine with interacting
binary stars or hole producing instabilities in expanding shock waves leading to illumination effects
from the central star that change the appearance of the nebula. 
Furthermore, there has been a lot of progress in the observation and
3D modeling of the kinematics, which is key to the understanding of the dynamics. 
Extensive observational catalogs are coming online for the kinematics, as well as some
very detailed proper motion measurements have been made. New techniques for 
morpho--kinematic 3D modeling help to make the interpretation of kinematic data more 
reliable and detailed. In addition to individual pointed observations, new surveys have lead
to the discovery of many PNe that show clear signs of interaction with the interstellar
medium. Systematic hydrodynamic models of the interaction have produced a general scheme for
the observed structure that results from the interaction of an evolving planetary nebula with 
the ISM. Detailed modeling of the dust-gas dynamics during the interaction with the ISM
have produced interesting predictions for future IR observations. Detailed models were worked out
for the structure of the bowshock and tail of Mira that was recently discovered in the UV.

\end{abstract}

\section{INTRODUCTION} 
In general terms the evolution of structure in planetary nebulae (PNe) appeared to have been 
pretty much solved with the interacting winds model by Kwok et al.(1978) and its generalization
(Kahn \& West, 1985; Balick 1987). A lot of theoretical work with numerical simulations 
during the nineties was able to 
reproduce findings from imaging and observations of the kinematics from spectroscopy 
(e.g. Mellema et al., 1991; Garc\'{\i}a-Segura \& L\'opez, 2000). 
From this illuminating research it seemed reasonable that the time scale for the structure formation 
was of the order of a thousand years or so. Hubble Space Telescope (HST) observations
of proto-planetary nebulae (PPNe) then made it clear that the complex structure must have been
formed at least an order of magnitude earlier and at a smaller spatial scale (Aaquist \& Kwok, 1996).
Therefore research trend is now going towards zooming in on nature of the ``central engine'', 
i.e. the central star(s) and the immediate surrounding on the scale of the size of an AGB star.

In this paper I shall {\em not} consider the work that has been done on the central engine
itself, but look at the research on the structure further out that might help to constrain the nature of 
the inner workings at scales that are spatially resolved with current imaging techniques. 
A key constraint for the dynamics, in addition to the morphology, is of course the kinematics.
Therefore, some emphasis will be put on the research that has been done in modeling observed
kinematics with hydrodynamic simulations as well as morpho--kinematic modeling, including 
new techniques that have been developed for the interpretation of observational data.\\

\begin{figure}[]
\begin{center}
 \includegraphics[width=5.3in]{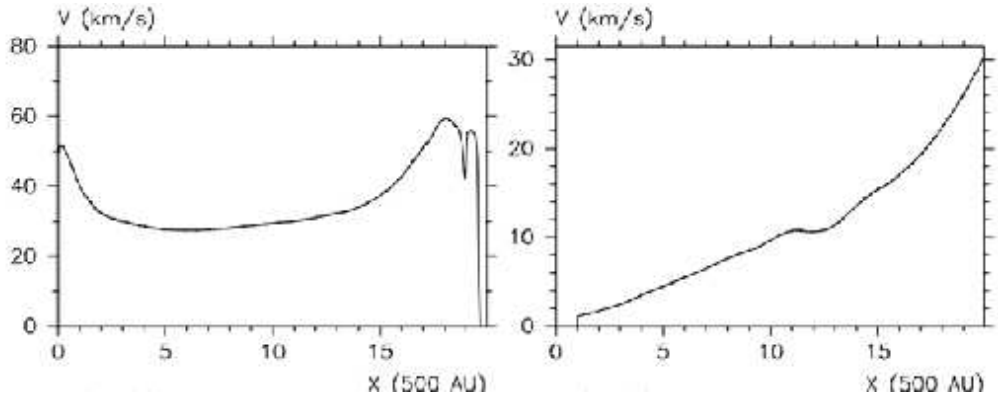}
 \caption{The weighted average velocity field produced by a jet (right) and a bullet as found by 
 Dennis et al., 2008). }
   \label{dennis}
\end{center}
\end{figure}

\section{JETS AND BULLETS} 
Jets and bullets have been suggested as the origin of the collimated features 
that can be observed in a number of young PNe and PPNe, such as He2--115 (Sahai \& Trauger, 1998) 
or CRL~618 (Trammel \& Goodrich, 2002). 
Raga et al. (2007) investigated the disintegration of bullets with high resolution 
3--D hydrodynamical simulations and find that the debris from the bullet assumes a velocity 
distribution that is roughly linear with distance to a reference position. Dennis et al. (2008)
looked at the shape and the velocity field of the bowshock or elongated lobes that bullets
and jets inflate during their propagation through the local environment. They find that the
lobes from continuous jets are ``thicker'', i.e. have a lower ratio between length and 
diameter, as compared to the bullets. Furthermore, the velocity profile of the lobes
is rather flat whereas that of the bullets shows a continuous nearly linear increase of 
velocity with distance from the source (see Fig.~\ref{dennis}). Their results suggest that it should be
possible to distinguish between an origin from bullets and jets using 
detailed observations of the kinematics of the lobes. 

Further research into the effects of knots and bullets on spectral line shapes and ratios was
done by Raga et al. (2008). They find that from the observed spectral 
line ratios it is hard or impossible to distinguish whether the structure is a stationary or 
slow object that is mainly photo--ionized or is a shock--ionized fast region. 
It is therefore necessary to include kinematic data to distinguish
between rather stationary photo--ionized structure and high-velocity shocked structures. 
Vel\'azquez et al. (2007) performed hydrodynamical simulations to model
 the properties of the water-maser source K3--35, finding that they were able to
reproduce the radio morphology and kinematics with a precessing jet that has a period of 
100~years and a semi--opening angle of 20 degrees. Guerrero et al. (2008) performed a detailed 
observational analysis of IC~4634 and included hydrodynamical simulations of precessing
variable jets to test their interpretation of the observations. In the past it has been rare to see
hydrodynamical simulations that were tailored for a particular object. With progress in
numerical computing capabilities, this is more viable now. Other examples for such simulations
are those presented by Monteiro \& Fal\c{c}eta--Gon\c{c}alves (2011; see also their 
poster in these proceedings) for NGC~40. They used to their dynamical simulations as input
to a detailed photoionization model. Vel\'azquez et al (2011) performed simulations for the 
Red Rectangle.

Akashi \& Soker (2008) did a detailed study on shaping PNe with light jets and partially collimated
winds. In Akashi et al. (2008) they compared the X-ray emission properties in PNe
that have been inflated by jets and partially collimated winds to investigate the problem 
of low observed X-ray temperatures. They find the hot shocked gas could enough through 
adiabatic expansion to explain the observations. However, for less collimated fast winds
require heat conduction to explain the low temperatures of the X-ray emitting gas. This is
agrees with earlier results from simulations of spherical winds by Steffen, M., et al. (2008). \\

\begin{figure}[]
\begin{center}
 \includegraphics[width=3.9in]{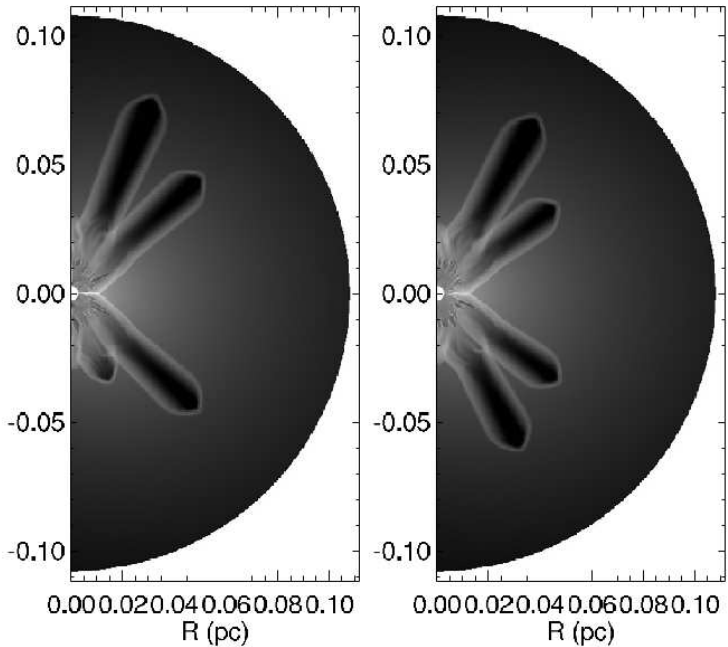}
 \caption{Cuts through the density distribution of hydrodynamical simulations of the 
 multi-polar structure formations through instabilities in photo--dissociation regions
 is shown (Garc\'{\i}a--Segura, 2011)}
   \label{ggs}
\end{center}
\end{figure}

\section{MULTI-POLAR LOBES} 
One of the most surprising discoveries about PPNe and PNe with HST--class high--resolution observations was the finding
of clearly multipolar objects such as CRL~618. The traditional interacting wind models are unable to explain this structure.
The first thought was then to introduce highly collimated ejections, like jets or bullets, that vary in direction and 
intensity over time (Sahai \& Trauger, 1998; Sahai, 2000). The time scale of this variability must be rather small, of the order or tens of years or so (see e.g. Riera et al., these proceedings).
Recently, Kwok (2010) proposed that bipolar and multipolar lobes of PNe could result from illumination and
ionization effects alone, without the need of collimated ejections. Based on this hypothesis, Garc\'{\i}a--Segura (2011) developed a multipolar structure forming model that is based on photo--{\bf dissociation} before the central star is hot enough to ionize its environment. This model explains multipolar features to appear during the PPN--phase and that may persist well into the evolved PN-phase (Fig.~\ref{ggs}). \\

\section{VELOCITY AND METALLICITY} 
The metallicity of ionized hot gas has a strong impact on its radiation and therefore also on its cooling properties. 
As a result, also the dynamical evolution will be influenced by the metallicity. Sch\"onberner et al. (2010)
have used spherically symmetric radiation-hydrodynamic simulations to investigate the impact of metallicity on the 
expansion of PNe. They find that the metallicity strongly changes the radial structure and velocity field. 
``The lower the metal content, the larger and thicker become the nebular shells and the smaller the wind--blown cavities'', 
which is, of course, similar to the difference between adiabatic (no radiative losses) and isothermal expansion (dynamical
heating is off-set by strong radiative losses). Their paper is very extensive and provides a detailed quantitative analysis
of the influence of metallicity on the structure and kinematics.

\begin{figure}[]
\begin{center}
 \includegraphics[width=5.3in]{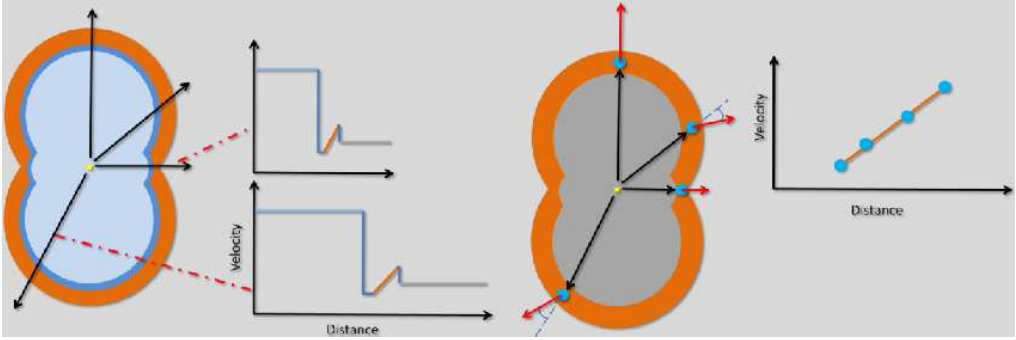}
 \caption{The radial velocity field in a PN is usually rather complex and is linear only over a small range (left). 
 Although the shape of the velocity field may be similar in different directions, quantitatively it will be different. 
 When considering a single ``shell'' (right), the plot of the speed versus distance will be closer to a linear increase, 
 although usually it still differs from a homologous expansion law, especially because of deviations from the radial direction.}
   \label{hubbleflow}
\end{center}
\end{figure}

\section{Morpho--kinematic modeling}
Since the advent of the HST, the observed complexity of PNe has increased dramatically and with it the potential of miss--classification due to orientation effects. Considerable effort has therefore been invested to improve 
the reconstruction of the 3--D structure from imaging and kinematic data (e.g. Sabbadin et al., 2004; Steffen \& L\'opez, 2006; Steffen et al., 2011; Hajian et al., 2007). Usually, a key ingredient to these reconstructions is the assumption of a
homologous expansion of the nebula, i.e. the velocity {\em vector} is proportional to position. Such a velocity field 
provides a linear mapping between the velocity (that can be measured with spatially resolved spectroscopy) and the position
along the line of sight. Assuming the existence of some symmetry -- such as sphericity or a cylindrical symmetry -- the 3--D structure can be reconstructed (including the inclination angle). This type of expansion is expected from a ballistic explosion. However, the GIW scenario implies a non-spherical continuously driven flow for a significant time of the nebular evolution, such that we can not expect a
ballistic flow. If we find that the velocity field is homologous with a precision that is inconsistent with the GIW model, 
then the driving force for the structure formation must have acted only for a very short period compared to the
current age of a nebula.

\begin{figure}[]
\begin{center}
 \includegraphics[width=4.4in]{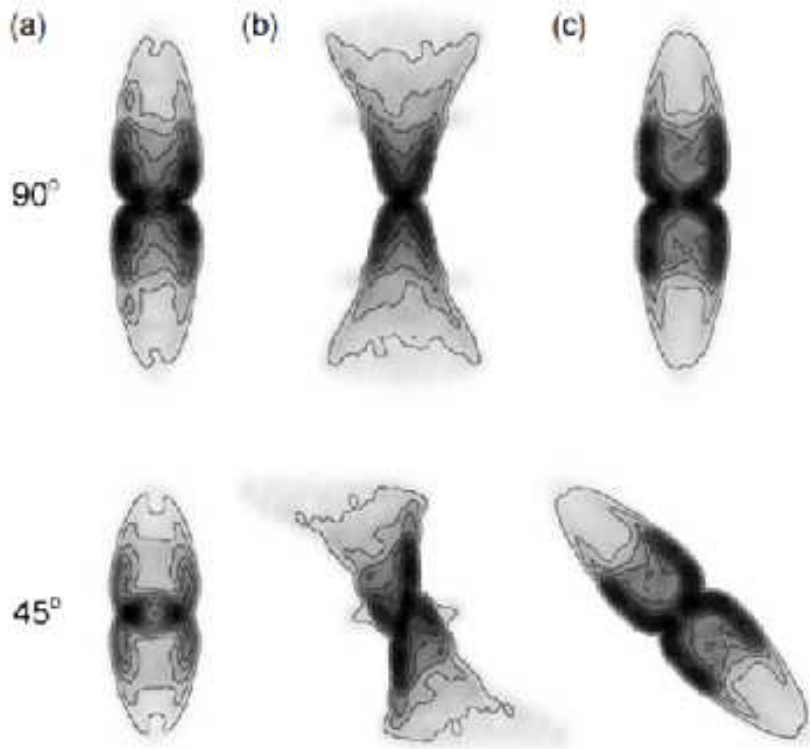}
 \caption{A hydrodynamic simulation of a cylindrically symmetric bipolar PN is show at two different viewing angles (col. a). Col. (b) is a reconstruction assuming a homologous expansion and as seen from a direction in the plane of the sky. In col. (c) the same view is shown but as reconstructed with the correct velocity law. The figure is from Steffen et al., 2009.}
   \label{reconstruction}
\end{center}
\end{figure}

Steffen et al. (2009) showed how the non-homologous velocity field may affect the reconstruction of structure along the line
of sight (Fig. \ref{reconstruction}) if a homologous expansion is assumed. They used hydrodynamical simulation as a know ''laboratory'' object for the reconstruction. The result depends on the inclination angle and can produce misclassifications. Fig. (\ref{reconstruction}) shows the same simulated object (col. a) and its reconstruction (col. b) for two different viewing angles. At 45 degrees the object might be falsely classified as point-symmetric. Column (c) shows the reconstruction when the correct velocity field is taken into account. The problem is, that either the velocity field has
to be assumed or a strong symmetry (like cylindrical symmetry) in order to have a unique reconstruction. Sometimes symmetric 
properties of parts of an object may help to solve ambiguities. The recent development of the interactive morpho-kinematic software SHAPE has contributed to a considerable increase of very detailed 3--D reconstructions of a variety of PNe (Steffen et al., 2011, see also http://www.astrosen.unam.mx/shape).

A key new kind of information to improve 3--D models of the structure and velocity field is beginning to appear from direct expansion and internal proper motion measurements. The time baseline of multi-epoch HST observations is now large enough to measure
the expansion component in the plane of the sky for the nearby objects (Li et al., 2002; Szyszka et al., 2011). They complement the spectroscopic Doppler-velocities that are already available (see e.g. L\'opez et al., these proceedings). The emerging proper motion measurements promise to produce a whole new game in morpho-kinematic modeling. New analysis techniques specific to proper motion measurements are already being developed in the form of criss-cross mapping and distance mapping (e.g. Steffen \& Koning, 2011; Akras \& Steffen, these proceedings). Radiation-hydrodynamic modeling to assess the impact of potential caveats for the interpretation of such data are needed, since there may be differences between the material flow as measured by the Doppler-effect and the pattern expansion near ionization fronts (Mellema 2004). Similarly, possible sources of systematic errors in the proper motion measurements should be evaluated, such as the overlapping of features from the front and back sides of a nebula.

\begin{figure}[]
\begin{center}
 \includegraphics[width=4.0in]{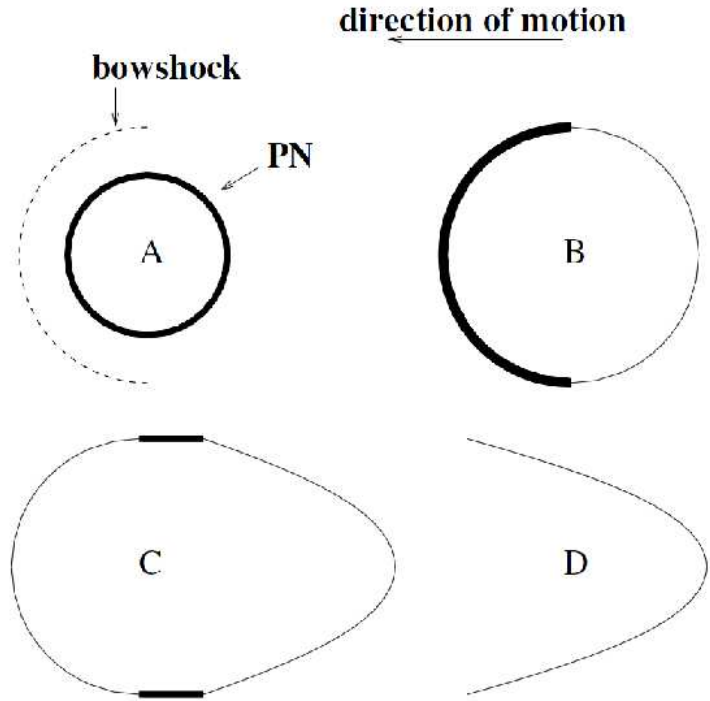}
 \caption{The classification scheme introduced by Wareing et al. (2007b) to describe the basic evolutionary stages of the interaction of PNe with the ISM. Small grains are on the left and large ones on the right.}
   \label{wareing}
\end{center}
\end{figure}

\section{ISM interaction}
\label{ISM}
The increased number of discoveries of low-surface brightness evolved PNe in surveys like IPHAS (Sabin et al., 2010) and MASH (Parker et al., 2006) has revealed a number of objects that show signs of interaction with the interstellar medium (ISM). Asymmetries seen in these objects have to be distinguished from intrinsic structure that has its origin in the formation process of the nebula on a small scale. After the initial work by Villaver et al. (2003, 2005), Wareing et al. (2007b) have conducted a series of simulations that systematically explores the main structural patterns in PNe that interact with the ISM. This has lead to a simple classification scheme that is based on the main, qualitatively different, phases of the interaction (Fig. \ref{wareing}).

Ueta et al. (2009) have devised a new method that uses the stellar wind interaction with the ISM to derive the 3-D ISM motion in the region near the star. They fit an analytical bowshock shape to the imaging observations in the IR and, with a few assumptions, obtain the velocity vector of the ISM in the star's neighbourhood.

\begin{figure}[]
\begin{center}
 \includegraphics[width=5.3in]{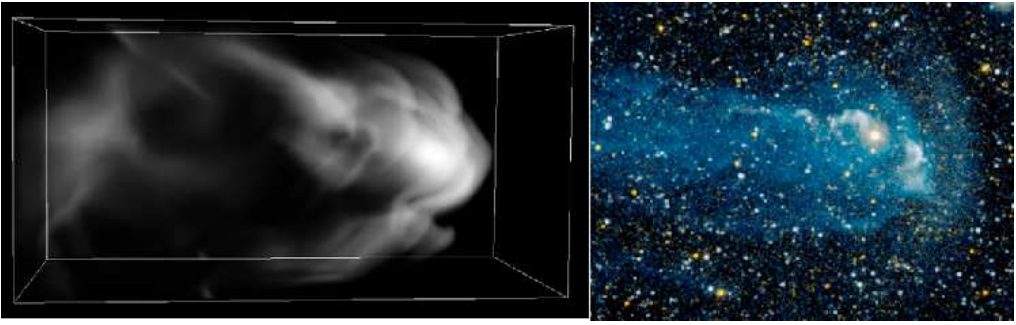}
 \caption{The bowshock around Mira has a very complex structure (right, GALEX, Martin et al., 2007). It was simulated by Raga et al. (2008) with a wind that has a latitude-dependent density structure and an axis that is oblique to the relative motion between the star and the ISM.}
   \label{mira}
\end{center}
\end{figure}

The spectacular tail of Mira that was detected in the UV (Martin et al., 2007) has also been the subject of analytical calculations and numerical simulations. Wareing et al. (2007a) came to the conclusion that the tail was approx. 470 thousand years old and that the rebrightening of the tail at some distance from the star was due to the star entering the Local Bubble (LB). Raga et al. explored the details of the bowshock structure (Fig. \ref{mira}) and the rebrightening with high resolution 2--D and 3--D simulations (Raga \& Cant\'o, 2008; Raga et al., 2008). They also investigated the effects of Mira entering the LB in Esquivel et al. (2010) and confirmed that the rebrightening of Mira's tail could be due to penetrating the LB.

Ransom et al. (2008) consider that the presence and deformation of the ISM magnetic field could explain the polarization of radio emission observed in the PN Sh--2 216. They propose that the motion of the PN through the ISM compresses and deforms the otherwise more uniform field. This results in enhanced and polarized radio emission towards the head of the bowshock that can also be observed in the optical.

The dynamical motion of dust in a bowshock of an AGB wind was explored in a series of simulations by van Marle et al. (2011, se also their poster contribution in these proceedings). They find that the small grains conform to the density distribution of the gas whereas the larger grains follow their inertia and can not be held back as easily by the gas. They predict that there should be an imprint of the separation of grain sizes in the IR radiation. This phenomenon would be the largest known selective dust filter in the universe.

\acknowledgements
The author acknowledges support by {\em Universidad Nacional Aut\'onoma de M\'exico} through grant DGAPA PAPIIT IN100401 and the {\em Alexander von Humboldt Stiftung}, as well as the support and the hospitality of the {\em Institut f\"ur Computergraphik} at the {\em Technische Universit\"at Braunschweig} during his sabbatical year.

\end{document}